\documentclass[12pt]{article}
\usepackage{epsf, cite, amssymb}
\usepackage{epsfig}
\usepackage{amsmath}
\makeatletter
\renewcommand\section{\@startsection {section}{1}{\z@}%
                                   {-3.5ex \@plus -1ex \@minus -.2ex}
                                   {2.3ex \@plus.2ex}%
                                   {\normalfont\large\bfseries}}
\renewcommand\subsection{\@startsection{subsection}{2}{\z@}%
                                     {-3.25ex\@plus -1ex \@minus -.2ex}%
                                     {1.5ex \@plus .2ex}%
                                     {\normalfont\bfseries}}
\makeatother

\let\non\nonumber

\let\s=\sigma

\newcommand{\bea}{\begin{eqnarray}}
\newcommand{\eea}{\end{eqnarray}}
\newcommand{\be}{\begin{equation}}
\newcommand{\ee}{\end{equation}}

\newcommand{\p}{\partial}

\newcommand{\C}[1]{$(\ref{#1})$}

\def\IZ{\relax\ifmmode\mathchoice
{\hbox{\cmss Z\kern-.4em Z}}{\hbox{\cmss Z\kern-.4em Z}}
{\lower.9pt\hbox{\cmsss Z\kern-.4em Z}} {\lower1.2pt\hbox{\cmsss
Z\kern-.4em Z}}\else{\cmss Z\kern-.4em Z}\fi}
\def\IR{\relax{\rm I\kern-.18em R}}

\def\one{{\hbox{ 1\kern-.8mm l}}}

\newlength{\bredde}
\def\slash#1{\settowidth{\bredde}{$#1$}\ifmmode\,\raisebox{.15ex}{/}
\hspace*{-\bredde} #1\else$\,\raisebox{.15ex}{/}\hspace*{-\bredde}
#1$\fi}

\newsavebox{\zzzbar}
\sbox{\zzzbar}
  {\setlength{\unitlength}{0.9em}
  \begin{picture}(0.6,0.7)
  \thinlines
  \put(0,0){\line(1,0){0.6}}
  \put(0,0.75){\line(1,0){0.575}}
  \multiput(0,0)(0.0125,0.025){30}{\rule{0.3pt}{0.3pt}}
  \multiput(0.2,0)(0.0125,0.025){30}{\rule{0.3pt}{0.3pt}}
  \put(0,0.75){\line(0,-1){0.15}}
  \put(0.015,0.75){\line(0,-1){0.1}}
  \put(0.03,0.75){\line(0,-1){0.075}}
  \put(0.045,0.75){\line(0,-1){0.05}}
  \put(0.05,0.75){\line(0,-1){0.025}}
  \put(0.6,0){\line(0,1){0.15}}
  \put(0.585,0){\line(0,1){0.1}}
  \put(0.57,0){\line(0,1){0.075}}
  \put(0.555,0){\line(0,1){0.05}}
  \put(0.55,0){\line(0,1){0.025}}
  \end{picture}}

\newcommand{\ena}{\end{eqnarray}}
\newcommand{\beqa}{\begin{eqnarray}}
\newcommand{\eeqa}{\end{eqnarray}}

\def\s{\sigma}

\begin{document}
\begin{titlepage}

\begin{center}

\today
\hfill         \phantom{xxx}

\vskip 2 cm
{\Large \bf Constraining the D3--brane effective action}\\
\vskip 1.25 cm { Anirban Basu\footnote{email address: abasu@ias.edu}$^{a}$ }\\
{\vskip 0.5cm $^{a}$ Institute for Advanced Study, Princeton, NJ 08540, USA\\}

\end{center}
\vskip 2 cm

\begin{abstract}
\baselineskip=18pt

We consider higher derivative corrections of the type $D^{2k} \mathcal{R}^2$ in the effective 
action of the D3--brane with trivial normal bundle. Based on the perturbative disc and annulus amplitudes,
and constraints of supersymmetry and duality, we argue that these interactions are protected, at least for small values 
of $k$. Their coefficient functions receive only a finite number of perturbative contributions, and non--perturbative 
contributions from D--instantons. We propose expressions for these modular forms for low values of $k$.    

\end{abstract}

\end{titlepage}

\pagestyle{plain}
\baselineskip=19pt

\section{Introduction}

Understanding the effective action of D--branes is important as it elucidates the role of 
non--perturbative duality symmetries in string theory. While a considerable amount of work has been done in analyzing
certain classes of protected interactions in the closed string sector in theories with sufficient supersymmetry, not much
is known about higher derivative corrections to the DBI action that describe the dynamics of D--branes at low 
energies. In this work, we consider certain interactions 
in the effective action of the self dual D3--brane in type IIB string theory with trivial normal bundle,
which preserves 16 supersymmetries. These interactions are of the 
form $D^{2k} \mathcal{R}^2$, where $k$ is a non--negative integer, 
which at the linearized level involve the two graviton scattering amplitude. 
In the Einstein frame, the coefficient functions 
of these interactions should be $SL(2,\mathbb{Z})$ invariant modular 
forms. These coefficient functions are expected to be protected at least for low values of $k$ 
for reasons mentioned below, and thus should receive perturbative 
contributions only upto a certain order in string perturbation theory, as well as non--perturbative contributions from
D--instantons. This is analogous
to the $D^{2k} \mathcal{R}^4$ interactions in the effective action of the closed string sector which preserve 32
supersymmetries.

Among the various higher derivative corrections in the effective action of the D3--brane, some interactions have been analyzed in the
literature~\cite{Bachas:1999um,Green:2000ke,Wyllard:2000qe,Fotopoulos:2001pt}. 
In particular, the $SL(2,\mathbb{Z})$ invariant coefficient function for the $\mathcal{R}^2$ interaction has been 
obtained in~\cite{Bachas:1999um}. The precise spacetime structure of the $\mathcal{R}^2$ interaction involves some ambiguities, as it
cannot be completely fixed using on-shell perturbative amplitudes~\cite{Fotopoulos:2001pt}. However, this will not concern us,
as we are interested in the coefficient functions of the $D^{2k} \mathcal{R}^2$ functions, and not the detailed spacetime
structure. Thus we define our interactions to be given at weak coupling at the linearized level by the two graviton disc amplitude with
boundary conditions appropriate to the D3--brane. Note that the analysis done in~\cite{Bachas:1999um} involved looking at the interaction
in the Wilsonian effective action, but we shall construct the modular forms by taking into account infrared effects as well, and constructing the 
duality invariant 1PI effective action. 

The $\mathcal{R}^2$ interaction receives perturbative contributions only from disc and 
annulus amplitudes~\cite{Bachas:1999um}, and non-perturbative contributions from D--instantons, and hence is protected. This suggests 
that the higher derivative corrections which are in the supermultiplet of the $D^{2k} \mathcal{R}^2$ interaction might also be protected, at least for 
sufficiently small values of $k$. One can motivate the non--renormalization property of the $\mathcal{R}^2$ supermultiplet by writing some of the
interactions as a superspace integral of a functional of a chiral superfield, over the eight broken chiral supersymmetries~\cite{Green:2000ke}. This
can be done starting from the on-shell linearized type IIB superspace of Howe and West~\cite{Howe:1983sra}, and imposing further constraints
due to the reduced supersymmetry. Thus, chirality of this construction makes the $\mathcal{R}^2$ multiplet protected. Now since F terms of this kind
can only involve chiral superspace integrals and there are 8 such fermionic integrals, by replacing a pair of chiral fermion bilinears by $\p_\mu$
and using dimensional analysis, it is conceivable that interactions of the form $D^{2k} \mathcal{R}^2$ are protected at least
for $k \leq 2$. This non--renormalization property should get extended to much higher values of $k$, and can presumably be analyzed along the lines 
of~\cite{Berkovits:2006vc} in the closed string sector.

Our aim is to argue about the existence of such non--renormalization theorems, based on constraints of the perturbative amplitudes,
supersymmetry, and $SL(2,\mathbb{Z})$ duality. We shall consider the $D^{2k} \mathcal{R}^2$ interactions for
$k=0, 1$ and $2$, and argue that they should receive only a finite number of perturbative contributions, as well as non--perturbative 
contributions from D--instantons. We first consider the low momentum expansion of the disc and annulus amplitudes
involving the scattering of two gravitons. The annulus amplitude contains terms logarithmic in the external momenta
coming from the propagation of massless modes in the loop. Converted to the Einstein frame, they lead to 
perturbative contributions to the modular forms which are logarithmic in the coupling. 
We next use constraints from supersymmetry to motivate that these
interactions should be protected, and use duality to construct expressions for them. In fact, our analysis
leads to the conclusion that these modular forms satisfy Poisson equations on moduli space. This should follow along 
the lines of~\cite{Green:1998by,Basu:2008cf} by arguments based on spacetime supersymmetry.
 
\section{The perturbative disc and annulus amplitudes}
    
We focus on the purely gravitational interactions in the D3--brane world volume of the form $D^{2k} \mathcal{R}^2$, 
which at the linearized level involve the two graviton amplitude. We shall only consider the case where 
the normal bundle is trivial. We consider the two leading perturbative contributions
which are given by the disc and the annulus amplitudes. We skip various technical details needed in 
calculating the amplitudes, for which we refer 
the reader to the literature. 

\subsection{The disc amplitude}

The disc amplitude involving the scattering of two gravitons with momenta $p_1$ and $p_2$ is given 
by~\cite{Garousi:1996ad,Hashimoto:1996bf}
\be \label{discamp}
\mathcal{A}_{\rm disc} = \tau_2 \frac{\Gamma (-\alpha' s/4) \Gamma (-\alpha' t/4)}{\Gamma (1
-\alpha' s/4 - \alpha' t/4 )} K(1,2), \ee
where $s= -2 p_1 \cdot D \cdot p_1$, $t= - 2 p_1 \cdot p_2$, and 
\be K(1,2) = -\frac{1}{2} (s a_1 - t a_2).\ee
Also $a_1$ and $a_2$ are given by
\bea a_1 &=& {\rm Tr} (\epsilon_1 \cdot D) p_1 \cdot \epsilon_2 \cdot p_1 - p_1 \cdot \epsilon_2 \cdot D \cdot \epsilon_1
\cdot p_2  -2 p_1 \cdot \epsilon_2 \cdot \epsilon_1 \cdot D \cdot p_1 \non \\ &&
- p_1 \cdot \epsilon_2 \cdot \epsilon_1 \cdot p_2 
- \frac{s}{4} {\rm Tr} (\epsilon_1 \cdot \epsilon_2)
+ (1 \leftrightarrow 2), \non \\ 
a_2 &=& {\rm Tr}(\epsilon_1 \cdot D) (2 p_1 \cdot \epsilon_2 \cdot D \cdot p_2 + p_2 \cdot D \cdot 
\epsilon_2 \cdot D \cdot p_2) \non \\ &&+ p_1 \cdot D \cdot \epsilon_1 \cdot D \cdot \epsilon_2 \cdot D \cdot p_2 
-p_2 \cdot D \cdot \epsilon_2 \cdot \epsilon_1 \cdot D \cdot p_1 \non \\ &&-\frac{s}{4} {\rm Tr} (\epsilon_1 \cdot 
D \cdot \epsilon_2 \cdot D ) +\frac{s}{4} {\rm Tr} (\epsilon_1 \cdot \epsilon_2)
\non \\ &&+\frac{1}{4} {\rm Tr} (\epsilon_1 \cdot D) {\rm Tr} (\epsilon_2 \cdot D) (s+t)+ (1 \leftrightarrow 2),\eea
where $\epsilon_i$ are the polarizations of the gravitons.
Momentum conservation along the world volume directions of the D3--brane is imposed by
\be \label{momcon}
p_1^\mu + (D \cdot p_1)^\mu  + p_2^\mu + (D \cdot p_2)^\mu =0, \ee
where the matrix $D^\mu_{~\nu}$ has components 1 along the world volume directions, and $-1$ along the transverse directions.
We perform the low momentum expansion of \C{discamp} using the relation
\be {\rm ln} \Gamma (1-x) = \gamma x + \sum_{n=2}^\infty \frac{\zeta (n) x^n}{n},\ee
which gives us that
\bea \label{expdisc}
\mathcal{A}_{\rm disc} &&= \frac{16 \tau_2}{\alpha'^2} \Big[ \frac{1}{st} - \zeta (2) \Big( \frac{\alpha'}{4} \Big)^2 - \zeta (3)
\Big( \frac{\alpha'}{4} \Big)^3 (s+t) \non \\ &&- \frac{1}{2} \Big( \frac{\alpha'}{4} \Big)^4 \Big( \zeta (4) (2 s^2 + 2 t^2 
+ 3st) - \zeta (2)^2 st \Big) + \ldots \Big] K(1,2).\eea
The first term in \C{expdisc} with poles in $s$ and $t$ gives the contribution from the DBI action as well as from
supergravity, while the others give contact interactions in the effective action which we schematically depict by $D^{2k}
\mathcal{R}^2$.  

\subsection{The annulus amplitude}

The leading correction to the disc amplitude is given by the annulus amplitude. This is given by~\cite{Pasquinucci:1997di,Lee:1997gwa} 
\bea \label{annamp}
\mathcal{A}_{\rm annulus} &&= \lambda \int_0^\infty \frac{dT}{T^3} \int_\Sigma d^2 z \int_\Sigma d^2 w  
e^{-\pi \alpha' s(z_2 - w_2)^2/2T} \times \non \\ && 
\Big\vert \frac{\theta_1 (z-w \vert iT)\theta_1 (\bar{z}- \bar{w} \vert iT)}{\theta_1 (\bar{z}+ w \vert iT) \theta_1
(z + \bar{w} \vert iT)} \Big\vert^{-\alpha' t/4} \Big\vert \frac{\theta_1 (z+\bar{z} \vert iT)\theta_1 
(w+ \bar{w} \vert iT)}{\theta_1 (\bar{z}+ w \vert iT) \theta_1(z + \bar{w} \vert iT)}
\Big\vert^{-\alpha' s/4} K(1,2),\eea
where $\lambda$ is a constant which can be fixed using unitarity. However, we shall fix it later using duality.
While integrating the vertex operators over the annulus worldsheet
$\Sigma$, we use the notation $z = x_1 + i y_1$, $w = x_2 + i y_2$, where we have to integrate over 
\be 0 \leq x_i \leq \frac{1}{2}, \quad  0 \leq y_i \leq T ,\ee
where $T$ is the modulus of the annulus.

We find it convenient to define $u = -2 p_1 \cdot D \cdot p_2$, and so mass shell constraints and momentum conservation 
\C{momcon} implies that
\be s + t + u =0 .\ee
For our purposes, it is more convenient to express \C{annamp} explicitly in terms of scalar propagators, which leads to
\bea \label{annamp2}
\mathcal{A}_{\rm annulus} &&= \lambda \int_0^\infty \frac{dT}{T} F(s,t,u;T)
K(1,2),\eea
where
\bea F(s,t,u;T) = \int_\Sigma \frac{d^2 z}{T} \int_\Sigma \frac{d^2 w}{T}  
{\rm exp}\Big[ \frac{\alpha' s}{2} \Big( \hat{P} (z+ \bar{z} \vert iT)+ \hat{P} (w + \bar{w}\vert iT)\Big) 
\non \\ + \frac{\alpha' t}{2} \Big( \hat{P} (z-w\vert iT)+ \hat{P} (\bar{z} -\bar{w}\vert iT)
\Big) +\frac{\alpha' u}{2} \Big( \hat{P} 
(z+ \bar{w}\vert iT)+ \hat{P} (\bar{z} + w\vert iT)\Big) \Big].\eea
We have used the relation 
\be \hat{P} (z \vert iT) = P (z \vert iT) - \frac{1}{2} {\rm ln} \vert \sqrt{2\pi} \eta (iT)\vert^2 ,\ee
where
\be P(z \vert iT) = -\frac{1}{4} {\rm ln} \Big\vert \frac{\theta_1 (z \vert iT)}{\theta_1'(0 \vert iT)} \Big\vert^2 
+ \frac{\pi z_2^2}{2 T}.\ee
Thus we have that~\cite{Green:1999pv}
\be \hat{P} (z \vert iT ) = \frac{1}{4\pi} \sum_{(m,n) \neq (0,0)} \frac{T}{m^2 + n^2 T^2} {\rm exp} \Big[ 
\frac{\pi}{T} \Big( \bar{z} (m + inT) - z (m-inT)\Big)  \Big] .\ee

We now have to evaluate \C{annamp2}. It contains two kinds of contributions: 

(i) analytic in the external momenta, and 

(ii) non--analytic in the external momenta, coming from the propagation of massless modes in the loop. 

The non--analytic
contributions to the amplitude are obtained both from the $T \rightarrow \infty$ and $T \rightarrow 0$ limits of the integral over the modulus.
While the non--analyticity in the $T \rightarrow \infty$ limit is from the massless open string modes in the loop, the
non--analyticity in the $T \rightarrow 0$ limit is interpreted as coming from the massless closed string modes in the dual
channel. 

Thus we split the integral over $T$ into three regimes: $[0,1/L_1], (1/L_1, L_2)$, and $[L_2, \infty)$, and consider the
limit $L_i \rightarrow \infty$ at the end. The regime $(1/L_1, L_2)$ gives the analytic contributions which are finite and independent of
$L_i$, as well as $L_i$ dependent contributions which diverge in this limit. The other two regimes $[0,1/L_1]$ and $[L_2, \infty)$ give 
the non--analytic contributions as well as $L_i$ dependent diverging ones. Of course, the total divergences cancel, giving the 
complete amplitude.

Thus we get that
\bea \label{annampanal}
\mathcal{A}_{\rm annulus}^{\rm analytic} &&= \lambda \int_{1/L_1}^{L_2} \frac{dT}{T} F(s,t,u;T)
K(1,2),\eea
and
\bea \label{annampnonanal}
\mathcal{A}_{\rm annulus}^{\rm nonanalytic} &&= \lambda \Big( \int_0^{1/L_1} + \int_{L_2}^\infty  \Big) \frac{dT}{T} F(s,t,u;T)
K(1,2).\eea
We first consider the analytic part of the annulus amplitude given by \C{annampanal}.    

\subsubsection{The analytic part of the annulus amplitude}

We consider the analytic part of the amplitude upto $O(\alpha'^2)$.
At $O(1)$, we get that
\be \label{ana0}
\mathcal{A}_{\rm annulus}^{\rm analytic} (O(1)) = \frac{\lambda}{4} K(1,2) ({\rm ln} L_2 + {\rm ln} L_1),\ee
while at $O(\alpha')$, we get that
\be \label{ana1}
\mathcal{A}_{\rm annulus}^{\rm analytic} (O(\alpha')) = \frac{\lambda \alpha' \zeta (2)}{8\pi} \Big( sL_2 + (t-u)\frac{L_1}{4})
\Big) K(1,2),\ee
where we have used the integrals \C{intone}.

At $O(\alpha'^2)$, we get that
\bea \label{anaO2}
\mathcal{A}_{\rm annulus}^{\rm analytic} (O(\alpha'^2)) &=& \frac{\lambda \alpha'^2}{8(4\pi)^2} \Big[ 
\zeta (2)^2 (s^2 + 2 t^2 + 2 u^2) \non \\ &&+ \zeta (2)^2 \Big( 4 s^2 {\rm ln} \Big( \frac{L_1}{\mu_1} \Big)
+ 2 s(s + 2t) {\rm ln} \mu_2  \Big) + f(L_1, L_2)\Big] K(1,2) ,\eea
where
\bea f(L_1, L_2) = \frac{\zeta (4)}{4} (2 s^2 + u^2 + 7 t^2) L_1^2 + 6 s^2 \zeta (4) L_2^2 ,\eea
and 
\bea {\rm ln} \mu_1 &=& -\frac{3 \zeta'(2)}{2 \zeta (2)} , \non \\ 
{\rm ln} \mu_2 &=& \frac{1}{40} {\rm ln} 2 + \frac{3\zeta'(4)}{8\zeta(4)}. \eea
To obtain \C{anaO2} we have used the integrals \C{inttwo}, \C{intthree}, \C{intfour}, \C{intfive}, \C{intsix}, and
\C{intseven}. 

Let us consider the logarithmically divergent terms in the limit $L_i \rightarrow \infty$ in \C{anaO2}.  Clearly, they must be cancelled by 
terms of the form ${\rm ln}(-\alpha' s L_i)$ coming from the non--analytic part of the amplitude, as we shall later see.
Thus we interpret the terms depending on ${\rm ln}\mu_i$ as the logarithmic scales  
in the amplitude rather than as the finite parts. Of course, there is no unambiguous way to do this, but this is natural based on
considerations of transcendality~\cite{Green:2008uj}. We shall later see that this is also natural based on constraints of duality.
Note that the ${\rm ln} \mu_2$ 
part of the expression does not contain any $L_i$ dependence, as the $L_2$ dependence cancels out on adding the various 
contributions. However we can always reinsert that dependence to write this as the logarithmic scale.       

\subsubsection{The non--analytic part of the annulus amplitude}

We next consider the non--analytic part of the amplitude given by \C{annampnonanal}. This can be rewritten as
\bea \mathcal{A}_{\rm annulus}^{\rm nonanalytic} &=&  \lambda \int_{L_1}^\infty \frac{dT}{T} \hat{F} (s,t,u;T) K(1,2) 
\non \\ &&+ \lambda \int_{L_2}^\infty  \frac{dT}{T} F(s,t,u;T) K(1,2) ,\eea
where
\bea \hat{F} (s,t,u;T) = \int_{\hat\Sigma} \frac{d^2 z}{T} \int_{\hat\Sigma} \frac{d^2 w}{T}  
{\rm exp}\Big[ \frac{\alpha' s}{2} \Big( \hat{P} (z- \bar{z} \vert iT)+ \hat{P} (w - \bar{w}\vert iT)\Big) 
\non \\ + \frac{\alpha' t}{2} \Big( \hat{P} (z-w\vert iT)+ \hat{P} (\bar{z} -\bar{w}\vert iT)
\Big) +\frac{\alpha' u}{2} \Big( \hat{P} 
(z- \bar{w}\vert iT)+ \hat{P} (\bar{z} -w \vert iT)\Big) \Big],\eea
where we have used
\be \hat{P} (-iz/T \vert i/T) = \hat{P} (z \vert iT). \ee
The domain of integration over $\hat\Sigma$ is
\be 0 \leq x_i \leq 1, \quad 0 \leq y_i \leq \frac{T}{2} .\ee
For the regime $[0,1/L_1]$, we find
it convenient to transform to variables appropriate to the dual 
closed string channel, such that the integration is over $[L_1, \infty)$ in the dual 
modulus. This has the advantage that we can use the standard asymptotic expansions of the propagators. 

We shall focus only on the logarithmically divergent contributions in the limit $L_i \rightarrow \infty$, 
as the polynomially diverging contributions trivially vanish in the sum.
So we need to consider the expression for the propagator for large $T$. This is given by~\cite{Green:1999pv,Green:2008uj}
\be \hat{P} (z | iT) = \hat{P}^\infty (z | iT) + \delta (z | iT), \ee
where
\be \hat{P}^\infty (z | iT) = \frac{T}{4\pi} \sum_{m \neq 0} \frac{1}{m^2} e^{2\pi i m y/T} = \frac{\pi T}{2}
\Big[ \Big(  \frac{y}{T} \Big)^2 - \frac{\vert y \vert}{T} + \frac{1}{6} \Big],\ee
which only depends on $y$ and is independent of $x$, and
\be \delta (z | iT) = \frac{1}{4} \sum_{m \neq 0} \frac{1}{\vert m \vert} e^{2\pi i m x - 2\pi \vert m y \vert },\ee
which is independent of the modulus $T$. The main idea is to treat the propagators perturbatively in $\delta (z | iT)$
about $\hat{P}^\infty (z | iT)$ and expanding to the required order in $\alpha'$~\cite{Green:1999pv,Green:2008uj}. 

First let us consider the contribution from the region $[L_2, \infty)$. Renaming $ y_1 
=\eta_1 T$ and $y_2 = \eta_2 T$, we get that
\bea \mathcal{A}_{{\rm annulus} L_2}^{\rm nonanalytic}  &=& 
\lambda K(1,2) \int_{L_2}^\infty \frac{dT}{T} \int_0^{1/2} d x_1 \int_0^{1/2}
d x_2 \int_0^1 d \eta_1  \int_0^1 d \eta_2 
e^{\frac{\alpha' \pi T s}{2} (\vert \eta_1 - \eta_2 \vert - (\eta_1 -\eta_2)^2)} \non \eea
\bea &&\times
{\rm exp} \Big[ \frac{\alpha' s}{2} \Big(\delta ( 2 x_1| iT) +\delta (2 x_2| iT) \non \\ &&
-\delta (x_1 + x_2 + i(\eta _1-\eta_2)T| iT) -\delta (x_1 + x_2 - i(\eta_1 -\eta_2)T| iT) \Big) 
\non \\ &&+ \frac{\alpha' t}{2} \Big(\delta (x_1 - x_2 + i(\eta_1 -\eta_2)T| iT)+\delta (x_1 - x_2 - i(\eta_1 -\eta_2)T| iT)
\non \\ && -\delta (x_1 + x_2 + i(\eta_1 -\eta_2)T| iT)-\delta (x_1 + x_2 - i(\eta_1 -\eta_2)T| iT) \Big)  \Big] .\eea

Considering the $O(1)$ contribution, we get\footnote{It is convenient to do the analysis setting $\eta_2 =0$ 
using translational invariance, and redefining $\eta_1 = \eta$.}
\bea \mathcal{A}_{{\rm annulus} L_2}^{\rm nonanalytic} (O(1)) = \frac{\lambda}{4} K(1,2) \int_0^1 d \eta 
E_1 \Big( -\frac{\pi \alpha' s L_2}{2} \eta (1 - \eta ) \Big) ,\eea
where $s < 0$\footnote{We shall evaluate integrals of this kind in the kinematically convergent regimes, and define them by
analytic continuation to the remaining regimes of the parameters. One should be able to precisely justify this statement
along the lines of~\cite{D'Hoker:1994yr}.}, and $E_1 (x)$ is the exponential integral.
To obtain the logarithmic contribution, we take the $s \rightarrow 0$ limit, to get that
\be \label{addmore}
\mathcal{A}_{{\rm annulus} L_2}^{\rm nonanalytic} (O(1)) = -\frac{\lambda}{4} 
{\rm ln} \Big( \frac{-\pi \alpha' s L_2}{\mu_3} \Big) K(1,2)+ \ldots,\ee
where we have used
\be E_1 (x) = - \gamma - {\rm ln} x + O(x ) , \ee
and
\be {\rm ln} \mu_3 = {\rm ln} 2 + 2 - \gamma .\ee

Next let us consider the contribution from the region $[L_1, \infty)$. Renaming $2 y_1 
=\eta_1 T$ and $2 y_2 = \eta_2 T$, we get that
\bea \mathcal{A}_{{\rm annulus} L_1}^{\rm nonanalytic}  &&= \frac{\lambda}{4} K(1,2) 
\int_{L_1}^\infty \frac{dT}{T} \int_{0}^{1} d x_1 \int_{0}^{1}
d x_2 \int_0^{1} d \eta_1   \int_0^{1} d \eta_2 
\non \\ && {\rm exp} \Big( \frac{\alpha' \pi T s}{8} (\eta_1 - \eta_2)^2 
+ \frac{\alpha' \pi T t}{4} (-2 \eta_1 \eta_2 + \eta_1 + \eta_2 -\vert \eta_1 -  \eta_2 \vert) \Big) \non \\ &&\times
{\rm exp} \Big[ \frac{\alpha' s}{2} \Big(\delta ( i \eta_1 T | iT) +\delta (i \eta_2 T | iT) \non \\ &&
-\delta (x_1 - x_2 + \frac{i}{2}(\eta_1 + \eta_2)T| iT) -\delta (x_1 - x_2 - \frac{i}{2}(\eta_1 +\eta_2)T| iT) \Big) 
\non \\ &&+ \frac{\alpha' t}{2} \Big(\delta (x_1 - x_2 + \frac{i}{2}(\eta_1 -\eta_2)T| iT)+\delta (x_1 - x_2 - \frac{i}{2}
(\eta_1 -\eta_2)T| iT)
\non \\ && -\delta (x_1 - x_2 + \frac{i}{2}(\eta_1 + \eta_2)T| iT)-\delta (x_1 - x_2 
- \frac{i}{2}(\eta_1 + \eta_2)T | iT) \Big)  \Big] .\eea

Again, considering the $O(1)$ contribution, we get that
\bea \mathcal{A}_{{\rm annulus} L_1}^{\rm nonanalytic} (O(1)) = \frac{\lambda}{2} K(1,2) \int_0^1 d \eta_1 \int_0^{\eta_1} d \eta_2
E_1 \Big( -\frac{\alpha' s \pi L_1}{8} (\eta_1 - \eta_2)^2 \non \\ - \frac{\alpha' t \pi L_1}{2} \eta_2 (1-\eta_1)\Big), \eea
for $s,t <0$. Thus taking the $s,t \rightarrow 0$ limit, we get that
\bea \label{nonanaL1}
\mathcal{A}_{{\rm annulus} L_1}^{\rm nonanalytic} (O(1)) = -\frac{\lambda}{4} \Big[ \gamma + {\rm ln} 
\Big( \frac{\pi \alpha' L_1}{2} \Big) \non \\ + 2 \int_0^1 d \eta_1 \int_0^{\eta_1} d \eta_2
{\rm ln} \Big( -\frac{s}{4} (\eta_1 - \eta_2)^2 - t \eta_2 (1-\eta_1) \Big) \Big] K(1,2) .\eea

The precise value of the integral in \C{nonanaL1} is not very relevant for our purposes, as the ${\rm ln}L_1$ dependence
has been extracted, and the conversion to the Einstein frame in order to analyze the implications of duality is also straightforward.
To evaluate the integral, we first integrate over $\eta_2$, after which for some of the terms that arise, we use the integral
representation
\be {\rm tan}^{-1} x = x \int_0^1 \frac{dt}{1 + x^2 t^2}, \ee
and do the $\eta_1$ integral. This gives us that
\bea \label{nonanaL11}
\mathcal{A}_{{\rm annulus} L_1}^{\rm nonanalytic} (O(1)) = -\frac{\lambda}{4} \Big[ {\rm ln} \Big( \frac{-\pi \alpha'
L_1 s}{\mu_4} \Big) + \sqrt{-\frac{t}{u}} \Big( {\rm tanh}^{-1} \sqrt{-\frac{u}{t}} \Big) \Big( \frac{4t}{s} -\frac{s}{u}
{\rm ln} (s/t) \Big) \non \\ + \frac{2t^2}{us} + \frac{4 t}{s} {\rm ln} 2 +\frac{t}{u}
{\rm ln} (s/t) +  \frac{s}{u} \sqrt{-\frac{t}{u}} \Big( Li_2 (\sqrt{-u/t}) -  Li_2 (-\sqrt{-u/t}) \Big) \non \\
+\frac{4s}{t} \sum_{n=0}^\infty \frac{(n+1)(n+2)}{(2n+3)} \Big( -\frac{u}{t} \Big)^n H_{n + 3/2} \Big] K(1,2) ,\eea
where
\be {\rm ln} \mu_4 = 3 {\rm ln} 2 + 3 -\gamma ,\ee
and 
\be H_{n + 3/2} = \gamma + \psi^{(0)} (n + 5/2)\ee
is the harmonic number. We have not evaluated the last term involving an infinite sum in \C{nonanaL11}.
There are various terms involving poles from supergravity and from the DBI action, as well as non--local terms
logarithmic in the external momenta. 

Thus, adding \C{ana0}, \C{addmore} and \C{nonanaL11}, we get that
\bea \label{O1}
\mathcal{A}_{{\rm annulus}} (O(1)) = -\frac{\lambda}{2} {\rm ln} \Big( \frac{-\pi \alpha' s}{\mu_5} \Big)  K(1,2)
-\frac{\lambda}{4} \Big[\sqrt{-\frac{t}{u}} \Big( {\rm tanh}^{-1} \sqrt{-\frac{u}{t}} \Big) \Big( \frac{4t}{s} -\frac{s}{u}
{\rm ln} (s/t) \Big) \non \\ + \frac{2t^2}{us} + \frac{4 t}{s} {\rm ln} 2 +\frac{t}{u}
{\rm ln} (s/t) +  \frac{s}{u} \sqrt{-\frac{t}{u}} \Big( Li_2 (\sqrt{-u/t}) -  Li_2 (-\sqrt{-u/t}) \Big) \non \\
+\frac{4s}{t} \sum_{n=0}^\infty \frac{(n+1)(n+2)}{(2n+3)} \Big( -\frac{u}{t} \Big)^n H_{n + 3/2} \Big] K(1,2) ,\eea
where
\be {\rm ln} \mu_5 = 2 {\rm ln} 2 + \frac{5}{2} -\gamma .\ee

For our purpose of understanding the constraints coming from $SL(2,\mathbb{Z})$ duality, we need to consider only the
first term in \C{O1}, as it gives a contribution to the modular form for the ${\mathcal{R}}^2$ interaction after converting to the
Einstein frame. From now on, we shall focus on only such terms. Thus
\bea \mathcal{A}_{{\rm annulus}} (O(1)) = -\frac{\lambda}{2} {\rm ln} \Big( \frac{-\pi \alpha' s}{\mu_5} \Big)  K(1,2)
+ \ldots. \eea

Next consider the relevant terms at $O(\alpha')$. From \C{ana1}, we see that there are no logarithmic divergences in $L_i$, thus
there cannot be terms of the form $s{\rm ln} (-\alpha' L_i s)$ from the non--analytic contribution. These are the terms which 
would have contributed to the modular form for the $D^2 \mathcal{R}^2$ interaction after converting to the Einstein frame. 
Thus the non--analytic contribution will cancel the linear divergence in $L_i$, and can also provide contributions of the type
$\alpha' s {\rm ln} (s/t)$ for example, which do not contribute to the modular form. Thus for our purposes
\bea A_{{\rm annulus}} (O(\alpha')) = 0 + \ldots . \eea 

Finally let us consider the contribution at $O(\alpha'^2)$. From \C{anaO2}, we see that the non--analytic contributions
must provide divergent contributions in $L_i$ which are power behaved, and cancel the $f(L_1, L_2)$ term, and it must also
provide logarithmically divergent contributions. Again, let us focus only on those which contribute to the modular form, i.e.,
the ones of the form $\alpha'^2 s^2 {\rm ln} (-\alpha' s)$. Without doing any calculation, from \C{anaO2}, it follows that
this contribution must be given by
\bea \label{gencon}
 A_{{\rm annulus}} (O(\alpha'^2)) = -\frac{\lambda \alpha'^2 \zeta (2)^2}{8(4\pi)^2} 
\Big[ 4 s^2 {\rm ln} \Big( \frac{-\alpha' \hat{p}^2 L_1}{\mu_p} \Big)
+ 2 s(s + 2t) {\rm ln} \mu_q  \Big] K(1,2)+ \ldots, \eea
where $\hat{p}^2$ is a scalar of order momentum squared (a linear combination of 
$s$ and $t$), and $\mu_p$ and $\mu_q$ are two scales. We interpret the ${\rm ln} \mu_i$ terms as scales
corresponding to ${\rm ln} L_i$ rather than as finite contributions, as finite 
contributions can only come from the analytic part of the integral over the modulus.  

Of course these scales can be determined by calculating the non--analytic 
contribution explicitly, which is not necessary for our purpose. Fixing them requires a detailed calculation of the logarithmic 
divergences, but this is not necessary to obtain the modular forms.    

However, to illustrate the appearance of a logarithmic term from the $[L_2, \infty)$ part of the integral
at this order, let us consider the simplest part of the integrals which gives such a term. We consider 
\bea \mathcal{A}_{{\rm annulus} L_2}^{\rm nonanalytic}  (O(\alpha'^2)) = \frac{\lambda \alpha'^2 K(1,2) s^2}{8} 
\int_{L_2}^\infty \frac{dT}{T} \int_0^{1/2} d x_1 \int_0^{1/2} d x_2 \int_0^1 d \eta   
\non \\ \times e^{\frac{\alpha' \pi T s}{2} \eta (1 - \eta) } 
\Big(\delta ( 2 x_1| iT) +\delta (2 x_2| iT) \Big)^2 + \ldots .\eea
Proceeding as before, this gives us
\bea \mathcal{A}_{{\rm annulus} L_2}^{\rm nonanalytic}  (O(\alpha'^2)) = -\frac{3\lambda \alpha'^2 \zeta (2)^2 s^2}{4 (4\pi)^2} 
{\rm ln} \Big( \frac{-\pi \alpha' s L_2}{\mu_3} \Big) K(1,2) + \ldots , \eea
for $s < 0$,
which yields a term of the form \C{gencon}. Thus the various such contributions must add up to determine ${\rm ln} \mu_q$
while the ${\rm ln} L_2$ term must cancel. 

\section{Constraints from supersymmetry and duality}

So how do these perturbative calculations constrain the modular forms associated with the various interactions? In the string frame,
the terms that are analytic in the external momenta
give rise to terms in the effective action of the form $D^{2k} \mathcal{R}^2$. The terms non--analytic in the external
momenta which are of the
form we considered above give rise to non--local terms in the effective action of the form ${\rm ln} (-\alpha' D^2) D^{2k} \mathcal{R}^2$.
On converting to the Einstein frame with metric $\hat{g}_{\mu\nu}$ defined by $g_{\mu\nu} = \tau_2^{-1/2} \hat{g}_{\mu\nu}$, the non--local 
terms give rise to local terms in the effective action with coefficients having a logarithmic dependence on $\tau_2$. Thus the local and 
non--local interactions in the string frame effective action both contribute to local interactions in the Einstein frame 
effective action, whose coefficients must be $SL(2,\mathbb{Z})$ invariant modular forms. In fact, for the cases we have looked at,
in the Einstein frame, these contributions are given by
\bea \label{Einsframe}
- \tau_2^{-1} \mathcal{A} &=& -\frac{16}{\alpha'^2 \hat{s} \hat{t}} \hat{\mathcal{R}}^2 + \Big( \zeta (2) 
\tau_2 + \frac{\lambda}{4} {\ln} \tau_2 \Big) \hat{\mathcal{R}}^2 - \frac{\zeta (3)}{4} \tau_2^{3/2} \alpha' \hat{u} 
\hat{\mathcal{R}}^2 \non \\ &&+ \frac{1}{16} \Big( \zeta (4) \tau_2^2 - \frac{\lambda}{16} \zeta (2) \tau_2 
\Big( 1 - \frac{2}{3} {\ln} \tau_2 \Big) \Big) \alpha'^2 \hat{s}^2 \hat{\mathcal{R}}^2  \non \\ &&
+ \frac{1}{16} \Big( \zeta (4) \tau_2^2 - \frac{\lambda}{12} \zeta (2) \tau_2 \Big)  \alpha'^2 \hat{t}^2 \hat{\mathcal{R}}^2 
\non \\ &&+  \frac{1}{64} \Big( \zeta (4) \tau_2^2 - \frac{\lambda}{3} \zeta (2) 
\tau_2 \Big) \alpha'^2 \hat{s} \hat{t} \hat{\mathcal{R}}^2 + \ldots , \eea
where we have set $K(1,2) = \mathcal{R}^2$, and ignored various irrelevant factors. We have absorbed a factor of $\tau_2^{-1}$
on the left hand side of \C{Einsframe} as that is cancelled by the same factor coming from converting $\sqrt{-g}$ to the Einstein 
frame, and thus drops out of the expression for the effective action. While the first term in \C{Einsframe} is from the DBI action 
and supergravity, the remaining terms have coefficients that must be $SL(2,\mathbb{Z})$ invariant modular forms.

To see the general structure, let us consider the constraints coming from supersymmetry. The invariance of the effective action
under supersymmetry leads to $\delta S =0$, and we can expand the supercharge as
\be \label{delsusy} \delta = \delta^{(0)} + \alpha'^2 \delta^{(2)} + \alpha'^3 \delta^{(3)} + \ldots, \ee 
and the effective action as
\be \label{delact} S = S^{(0)} + \alpha'^2 S^{(2)} + \alpha'^3 S^{(3)} + \ldots.\ee
Considering these constraints at $O(\alpha'^k)$, $\delta^{(0)} S^{(k)}$ and $\delta^{(k)} S^{(0)}$ should produce 
Laplace equation on moduli space, while the contributions from terms at intermediate orders in $\alpha'$ are expected to give source 
terms. One should be able to directly deduce these equations for interactions in the effective action of the form $\hat{G}^{2k}
\Lambda^8$, where $\hat{G}$ is the supercovariant 3--form~\cite{Schwarz:1983qr}, and $\Lambda$ is the gaugino, 
along the lines of~\cite{Basu:2008cf}.  

We expect nice equations at least for small values of $k$, because the theory preserves 16 supersymmetries, and also because it
contains maximally supersymmetric Yang--Mills
in its non--gravitational sector. As before, we schematically depict the terms of relevance in the effective action as 
$\hat{D}^{2k} \hat{\mathcal{R}}^2$.
From \C{delsusy} and \C{delact} it follows that the $\hat{\mathcal{R}}^2$ and the $\hat{D}^2 \hat{\mathcal{R}}^2$ interactions 
do not have source terms, while all the others do. Thus for the $\hat{\mathcal{R}}^2$ interaction, we have a term in the effective action
given by\footnote{We choose the $\alpha'$ dependence such that the DBI action is $O(1)$ in $\alpha'$.}   
\be \alpha'^2 \int d^4 x \sqrt{-\hat{g}} f_1 (\tau, \bar\tau) \hat{\mathcal{R}}^2 ,\ee
where $f_1 (\tau, \bar\tau)$ is given by
\be \label{f1val}
f_1 (\tau, \bar\tau) = 2 \zeta (2) \tau_2 + \frac{\lambda}{2} {\rm ln} \tau_2 + \ldots. \ee
Because there are no sources, this should satisfy Laplace equation on moduli space, which is solved by a specific 
non--holomorphic Eisenstein series, fixed entirely by the disc amplitude in \C{f1val}.  
Thus $f_1 (\tau, \bar\tau)$ is given by~\cite{Bachas:1999um} 
\bea f_1 (\tau, \bar\tau) = E_1 (\tau, \bar\tau) &=& 2 \zeta (2) \tau_2 -\pi {\rm ln} \tau_2 \non \\ && + 2\pi \sqrt{\tau_2}
\sum_{m \neq 0, n \neq 0} \Big\vert \frac{m}{n} \Big\vert^{1/2} K_{1/2} (2\pi \vert mn \vert \tau_2 )
e^{2\pi i mn \tau_1},\eea
which also sets
\be \lambda = -2\pi. \ee
The expression for this modular form has also been argued directly in~\cite{Bachas:1999um} by 
computing the one loop two graviton scattering
amplitude in eleven dimensional supergravity compactified on $T^2$ in the world volume theory of the M5--brane wrapping the $T^2$. The disc 
amplitude and the D--instanton contributions were explicitly calculated, while the annulus amplitude was not, as it 
does not contribute to the Wilsonian effective action. We complete the annulus amplitude calculation in appendix B
to show the origin of this term from supergravity.

The next term in \C{Einsframe} leads to the interaction  
\be \alpha'^3 \int d^4 x \sqrt{-\hat{g}} f_2 (\tau, \bar\tau) \hat{D}^2 \hat{\mathcal{R}}^2 ,\ee
where $f_2 (\tau, \bar\tau)$ is given by
\be f_2 (\tau, \bar\tau) = 2 \zeta (3) \tau_2^{3/2} + \ldots, \ee
while the $\tau_2^{1/2}$ term coming from the annulus amplitude vanishes. Again the absence of sources fixes
$f_2 (\tau, \bar\tau)$ to be given by 
\bea f_1 (\tau, \bar\tau) = E_{3/2} (\tau, \bar\tau) &=& 2 \zeta (3) \tau_2^{3/2} + 4 \zeta (2) \tau_2^{-1/2} \non \\ &&
+ 4\pi \sqrt{\tau_2} \sum_{m \neq 0, n \neq 0} \Big\vert \frac{m}{n} \Big\vert K_1 (2\pi \vert mn \vert \tau_2 )
e^{2\pi i mn \tau_1} .\eea
This receives the leading perturbative contribution from the disc, and the subleading contribution from the sum of the
pants diagram and the torus with a disc removed.

The final three terms in \C{Einsframe} lead to an interaction schematically depicted by
\be \alpha'^4 \int d^4 x \sqrt{-\hat{g}} f_3 (\tau, \bar\tau) \hat{D}^4 \hat{\mathcal{R}}^2 ,\ee
where there are three such interactions corresponding to the three independent spacetime structures $\hat{s}^2 \hat{\mathcal{R}}^2, 
\hat{t}^2 \hat{\mathcal{R}}^2$ and $\hat{s} \hat{t} \hat{\mathcal{R}}^2$. We denote the corresponding modular forms as $F_i (\tau, \bar\tau)$ respectively
for $i=1,2,3$. This is the first case where source terms can arise in the expressions for these equations. From \C{Einsframe}
and from constraints of supersymmetry, it is natural to expect that each $F_i$ can split into a sum of modular forms 
\be F_i (\tau, \bar\tau) = \sum_a f_{ia} (\tau, \bar\tau) ,\ee
each of which satisfies
\be 4 \tau_2^2 \frac{\p^2}{\p \tau \p \bar\tau} f_{ia} (\tau, \bar\tau) = \lambda_{1ia} f_{ia} (\tau, \bar\tau)
+ \lambda_{2ia} \Big( E_1 (\tau, \bar\tau) \Big)^2 .\ee 
For the modular forms $F_2$ and $F_3$, the 
coefficient of the ${\rm ln} \tau_2$ contribution at one loop must vanish after summing over all the contributions. It also leads
to the prediction that the next perturbative contribution given by the sum of the pants diagram, and the torus with a disc removed,
can be non--vanishing, and can receive a contribution from the source term proportional to $({\rm ln} \tau_2)^2$. 
This type of Poisson equations have arisen in ten and nine dimensions in the bulk theory~\cite{Green:2005ba,Basu:2007ck,Basu:2008cf}, 
while the degeneracy of modular forms has been observed in~\cite{Green:2008bf}.

Because of the large amount of supersymmetry, this structure could possibly persist for higher $k$. For example, consider the 
next case of the $\hat{D}^6 \hat{\mathcal{R}}^2$ interaction, which has
two independent spacetime structures for the disc amplitude given by $\tau_2^{5/2} \zeta (2) \zeta (3) \hat{s} \hat{t} \hat{u}$ 
and $\tau_2^{5/2} \zeta (5) \hat{u} (\hat{s}^2 + \hat{t}^2 +\hat{u}^2)$. While the former could give a modular form $G (\tau, \tau) 
= \sum_{a} g_a (\tau, \bar\tau)$, where $g_a$ satisfies
\be 4 \tau_2^2 \frac{\p^2}{\p \tau \p \bar\tau} g_a (\tau, \bar\tau) = \s_{1a} g_a (\tau, \bar\tau)
+ \s_{2a} E_1 (\tau, \bar\tau) E_{3/2} (\tau, \bar\tau) ,\ee
the later could give the modular form $E_{5/2} (\tau, \bar\tau)$.   

For values of $k$ where this pattern holds, the general structure of these 
equations shows that the $D^{2k} \mathcal{R}^2$ interactions receive only a finite
number of perturbative contributions, which can be argued recursively. This follows because for low $k$, this is the case, 
and these modular forms act as sources for the ones at higher $k$, and also the Laplace equation part of the equation
can give at most two perturbative contributions.

\section*{Acknowledgements}

The work of A.~B. is supported in part by NSF Grant No.~PHY-0503584 and the William D. Loughlin membership.

\appendix

\section{Relevant summations and integrals}

In evaluating the analytic part of the annulus amplitude, we need various summations and integrals which we mention below. 

At $O(\alpha')$, we need the integrals 

\bea \label{intone}
&&\int_{1/L_1}^{L_2} \frac{dT}{T} \int_\Sigma \frac{d^2 z}{T} \hat{P} (z + \bar{z} \vert iT) = \frac{\zeta (2) L_2}{4\pi},\non \\
&&\int_{1/L_1}^{L_2} \frac{dT}{T} 
\int_\Sigma \frac{d^2 z}{T} \int_\Sigma \frac{d^2 w}{T} \hat{P} (z -w \vert iT) \non \\ &&
= - \int_{1/L_1}^{L_2} \frac{dT}{T} \int_\Sigma \frac{d^2 z}{T} \int_\Sigma \frac{d^2 w}{T} \hat{P} (z +\bar{w} \vert iT) 
\non \\ &&=\frac{\zeta (2) L_1}{32\pi}, \eea
where we have used
\be  \sum_{n \neq 0} \Big( 1 - (-1)^n \Big) \frac{1}{n^4}= \frac{\pi^2 \zeta (2)}{4} .\ee

At $O(\alpha'^2)$, we need more complicated integrals which we list below.  

\bea \label{inttwo} &&\int_{1/L_1}^{L_2} \frac{dT}{T} \int_\Sigma \frac{d^2 z}{T} \hat{P} (z + \bar{z} \vert iT)^2 \non \\ 
&&= \frac{1}{2(4\pi)^2} \sum_{(m,n) \neq (0,0), (p,n) \neq (0,0)} \int_{1/L_1}^{L_2} dT
\frac{T}{(m^2 + n^2 T^2)(p^2 + n^2 T^2)}\non \\ &&= \frac{1}{2 (4\pi)^2} 
\Big[\zeta (4) L_1^2 + 2 \zeta (2)^2 L_2^2 + 2 \zeta (2)^2 + 8 \zeta (2)^2 
{\rm ln} \Big( \frac{L_1}{\mu_1} \Big) \Big],\eea
where
\be {\rm ln} \mu_1 = -\frac{3 \zeta'(2)}{2 \zeta (2)},\ee
and we have used
\be \sum_{k>0, k \neq m} \frac{1}{k^2 - m^2} = \frac{3}{4m^2},\ee
and 
\be \zeta' (s) = -\sum_{m=1}^\infty \frac{{\rm ln}m}{m^s}.\ee

\bea \label{intthree} \int_{1/L_1}^{L_2} \frac{dT}{T} \int_\Sigma \frac{d^2 z}{T} \int_\Sigma \frac{d^2 w}{T}
\hat{P} (z + \bar{z} \vert iT) \hat{P} (w + \bar{w} \vert iT) = \frac{\zeta (2)^2 L_2^2}{2(4\pi)^2}. \eea

\bea \label{intfour} && \int_{1/L_1}^{L_2} \frac{dT}{T} \int_\Sigma \frac{d^2 z}{T} \int_\Sigma \frac{d^2 w}{T} 
\hat{P} (z -w \vert iT)^2 \non \\ &&= \int_{1/L_1}^{L_2} \frac{dT}{T} \int_\Sigma \frac{d^2 z}{T} \int_\Sigma \frac{d^2 w}{T} 
\hat{P} (z -w \vert iT) \hat{P} (\bar{z} -\bar{w} \vert iT)  \non \\ &&= \frac{1}{(4\pi)^2} \int_{1/L_1}^{L_2} \frac{dT}{T}
\Big[ \frac{1}{4} \sum_{(m,n) \neq (0,0)} \frac{T^2}{(m^2 + n^2 T^2)^2} \non \\ &&+  2\sum_{(m,n) \neq (0,0) , (m,q) \neq (0,0) , n \neq q}
\frac{(1-(-1)^{n-q}) T^2}{(2\pi)^2 (n-q)^2 (m^2 + n^2 T^2) (m^2 + q^2 T^2)}\Big] \non \\ &&= \frac{1}{(4\pi)^2} 
\Big[\frac{5 \zeta (4)}{16} L_1^2 + \frac{\zeta (4)}{4} L_2^2 + \frac{\zeta (2)^2}{2} + \frac{\zeta (2)^2}{2} {\rm ln} \Big( \frac{L_2}{\mu_2} \Big)  
\Big],\eea
where
\be {\rm ln} \mu_2 = \frac{1}{40} {\rm ln} 2 + \frac{3\zeta'(4)}{8\zeta(4)},\ee
and we have used
\bea \sum_{m=0}^\infty \frac{{\rm ln} (2m+1)}{(2m+1)^s} = -{\rm ln} 2 \frac{\zeta (s)}{2^s} - \Big( 1 -\frac{1}{2^s}
\Big) \zeta' (s) .\eea
To obtain the above results, we made use of the summations
\bea && \sum_{m=1}^\infty \frac{1}{(2m+1)^2 (2m -2n +1)^2} = 
\frac{2 \psi^{(0)} (\frac{3}{2} -n) + n \psi^{(1)} (\frac{3}{2} -n)}{16 n^3} \non \\ 
&&~~~~~~~~~~~~~~~~~~~~~~~~~~~~~~~~~~~~~~~+ \frac{-8 +  4 \gamma -8n + \pi^2 n + 4 {\rm ln} 4}{32 n^3}, \non \\ &&
\sum_{m=1}^\infty \frac{1}{(2m-1)^2 (2m +2n -1)^2} = 
\frac{-2 \psi^{(0)} (\frac{1}{2} +n) + n \psi^{(1)} (\frac{1}{2} +n)}{16 n^3} \non \\ &&~~~~~~~~~~~~~~~~~~~~~~~~~~~~~~~~~~~~~~~+
\frac{-4 \gamma + \pi^2 n - 4 {\rm ln} 4}{32 n^3}, \non  \eea
\bea
&&\sum_{m \neq 0} \frac{3m-1}{m^5 (m - 1/2)^2} = -8 \zeta (4) + 128 \zeta (2) -192, \non \\
&&\sum_{m \neq 0} \frac{1}{m^2 (2 m - 1)^2} = 8 \zeta (2) -12, \non \\
&&\sum_{l=1}^\infty \frac{1}{[2l-(2m+1)]^2 [(2m+1)^2 - (2l)^2]} = -\frac{1}{8(2m+1)^4} \non \\ 
&& + \frac{2 \psi^{(1)} (1/2 -m) + (2m+1)
\psi^{(2)} (1/2 -m)}{32 (2m+1)^2},  \non \\ 
&&\sum_{l=1}^\infty \frac{1}{[2l+(2m+1)]^2 [(2m+1)^2 - (2l)^2]} = -\frac{1}{8(2m+1)^4} \non \\ &&+\frac{2 \psi^{(1)} (3/2 +m) - (2m+1)
\psi^{(2)} (3/2 +m)}{32 (2m+1)^2} ,\non \\ &&\sum_m
\frac{1}{[(2m+1) - 2l]^2 [(2l)^2 - (2m+1)^2]} = \frac{\pi^2}{64 l^2}.\eea
which leads to
\bea &&\sum_{m \neq 0, n \neq 0} \frac{1- (-1)^n}{m^2 n^4} ( {\rm ln} n^2 - {\rm ln} m^2)= \zeta (2) \Big( \pi^2 \zeta' (2) - \zeta (4) {\rm ln} 2
- 15 \zeta' (4) \Big), \non \\ &&\sum_{n \neq 0, q \neq 0, n \neq q} \frac{1- (-1)^{n-q}}{n^2 q^2 (q-n)^2} = \frac{\pi^2 \zeta (4)}{4} , 
\non \\  &&\sum_{n \neq 0, q \neq 0, n \neq q} \frac{(1-(-1)^{n-q})({\rm ln} 
q^2 - {\rm ln} n^2)}{(n-q)^2 (q^2 - n^2)} = - \pi^2 \zeta'(2) + \frac{5}{8} \zeta (4) {\rm ln} 2
+\frac{75}{8} \zeta'(4) .\eea

Here $\psi^{(n)}$ is the Polygamma function, and we also make use of the recurrence formula
\be \label{recur}
\psi^{(n)} (z +1) - \psi^{(n)} (z) = (-1)^n n! z^{-n-1}, \ee
and the reflection formula
\be \label{reflect}
\psi^{(n)} (1-z) + (-1)^{n+1} \psi^{(n)} (z) = (-1)^n \pi \frac{d^n}{dz^n} {\rm cot} \pi z .\ee

\bea \label{intfive} &&\int_{1/L_1}^{L_2} \frac{dT}{T} \int_\Sigma \frac{d^2 z}{T} \int_\Sigma \frac{d^2 w}{T} 
\hat{P} (z + \bar{w} \vert iT)^2 \non \\  &&= \int_{1/L_1}^{L_2} \frac{dT}{T} \int_\Sigma \frac{d^2 z}{T} \int_\Sigma \frac{d^2 w}{T} 
\hat{P} (z +\bar{w} \vert iT) \hat{P} (\bar{z} + w \vert iT)  \non \eea
\bea
&&= \frac{1}{(4\pi)^2} \int_{1/L_1}^{L_2} \frac{dT}{T}
\Big[ \frac{1}{4} \sum_{(m,n) \neq (0,0)} \frac{T^2}{(m^2 + n^2 T^2)^2} \non \\ &&-  2\sum_{(m,n) \neq (0,0) , (m,q) \neq (0,0) , n \neq q}
\frac{(1-(-1)^{n-q}) T^2}{(2\pi)^2 (n-q)^2 (m^2 + n^2 T^2) (m^2 + q^2 T^2)}\Big] \non \\ &&= \frac{1}{(4\pi)^2} 
\Big[ \frac{3 \zeta (4)}{16} L_1^2 + \frac{\zeta (4)}{4} L_2^2 + \frac{\zeta (2)^2}{2} - \frac{\zeta (2)^2}{2} {\rm ln} \Big( \frac{L_2}{\mu_2} \Big)  
\Big].\eea

\bea \label{intsix} &&\int_{1/L_1}^{L_2} \frac{dT}{T} \int_\Sigma \frac{d^2 z}{T} \int_\Sigma \frac{d^2 w}{T} 
\hat{P} (z + \bar{z} \vert iT) \hat{P} (z -w \vert iT) \non \\ &&= -\int_{1/L_1}^{L_2} \frac{dT}{T} 
\int_\Sigma \frac{d^2 z}{T} \int_\Sigma \frac{d^2 w}{T} 
\hat{P} (z + \bar{z} \vert iT) \hat{P} (z +\bar{w} \vert iT) \non \\ &&= \frac{1}{2 (2\pi)^4} \int_{1/L_1}^{L_2} \frac{dT}{T}
\sum_{q \neq 0, (m,n) \neq (0,0)} \frac{1-(-1)^q}{q^3 (2n +q) (m^2 + n^2 T^2) }\non \\ &&=
\frac{1}{4(4\pi)^2} \Big[ -\frac{\zeta(4)}{4}
L_1^2 + \zeta(2)^2 {\rm ln} L_2 \Big], \eea
where we have used
\bea \sum_{m=1}^\infty \frac{1-(-1)^m}{m^3 (2n +m)}  = \frac{1}{16n^3} \Big( \psi^{(0)} (n + \frac{1}{2}) 
- \psi^{(0)} (n+1) + 2 \psi^{(0)} (2n +1) \Big) \non 
\\ + \frac{2\gamma -n\pi^2 +{\rm ln}4 + 14 n^2 \zeta (3)}{16n^3}, \eea
the recurrence and reflection relations \C{recur} and \C{reflect}, as well as the doubling relation
\be 2 \psi^{(0)} (2z) = \psi^{(0)} (z) + \psi^{(0)} (z + \frac{1}{2}) + 2 {\rm ln} 2 .\ee
This gives us that
\bea \sum_{m \neq 0, n \neq 0} \frac{1-(-1)^m}{m^3 (2n +m)} = -\frac{\pi^2}{4} \zeta (2), \non \\
\sum_{m \neq 0, n \neq 0} \frac{1-(-1)^m}{m^3 n^2 (2n +m)} = -\frac{\pi^2}{4} \zeta (4),\non \\
\sum_{m \neq 0, n \neq 0} \frac{1-(-1)^m}{m^3  (2n +m)} {\rm ln} n^2 = \frac{\pi^2}{2} \zeta'(2).\eea

\bea \label{intseven} \int_{1/L_1}^{L_2} \frac{dT}{T} \int_\Sigma \frac{d^2 z}{T} \int_\Sigma \frac{d^2 w}{T} 
\hat{P} (z -w \vert iT) \hat{P} (z +\bar{w} \vert iT) = \frac{\zeta (4) L_2^2}{4 (4\pi)^2}.\eea

\section{The $\mathcal{R}^2$ annulus amplitude from supergravity}

Dropping irrelevant factors, the one loop two graviton amplitude is given by~\cite{Bachas:1999um}
\bea \label{sugra}
V A^{\rm sugra} =  2\zeta (2) \tau_2 + 2\pi \sqrt{\tau_2} \sum_{m \neq 0, n \neq 0} \Big\vert \frac{m}{n} 
\Big\vert^{1/2} K_{1/2} (2\pi \vert mn \vert \tau_2) e^{2\pi i mn \tau_1} \non \\
+ \pi \sqrt{\tau_2 V} \int_0^\infty \frac{dt}{t^{3/2}} \sum_{m \neq 0} e^{-\pi V m^2 \tau_2/t} .\eea

In \C{sugra}, $V$ is the volume of $T^2$ in the M theory metric, and $\tau$ is the complex structure. In taking the type IIB limit,
we have to take $V \rightarrow 0$, while keeping $\tau$ fixed, and identifying it with the type IIB complexified coupling. The first
two terms in \C{sugra} give the disc amplitude and the D--instanton contributions respectively, and we calculate the remaining term.
This is easily done using a calculation already done in~\cite{Green:2008uj}, in a completely different context.   
We write
\be \label{doneint}
I = \int_0^\infty \frac{dt}{t^{3/2}} \sum_{m \neq 0} e^{-\pi V m^2 \tau_2/t} = -2 + \int_1^\infty \frac{dt}{\sqrt{t}}
\sum_{m \neq 0} e^{-\pi V m^2 \tau_2 t} + \frac{1}{\sqrt{\tau_2 V}} \int_0^1 \frac{dt}{t} \sum_{m \neq 0} e^{-\pi m^2/(V \tau_2 t)}.\ee
In obtaining \C{doneint}, we first split the integral over $t$ from $(0, \infty)$ into two integrals from $(0,1)$ and $(1,\infty)$,
and perform Poisson resummation on the integrand in the second integral. This integral has been already done in (5.23) and (5.24) of~\cite{Green:2008uj} 
(we set $r^2 = 1/(V \tau_2)$), and gives us that
\be I = -\frac{1}{\sqrt{V \tau_2}} \Big( {\rm ln} (V \tau_2 ) - \gamma +{\rm ln} 4\pi\Big) ,\ee
leading to
\bea \label{fullsugra}
V A^{\rm sugra} =  E_1 (\tau, \bar\tau) -\pi{\rm ln}  (4\pi V e^{-\gamma} ).\eea
Now taking the type IIB limit, the last term gives a logarithmic divergence which 
can be renormalized by adding a suitable counterterm. It can give a finite $\tau$ independent contribution
to $V A^{\rm sugra}$ as in~\cite{Dixon:1990pc}.


\providecommand{\href}[2]{#2}\begingroup\raggedright\endgroup

\end{document}